\documentclass[twocolumn,aps,prb]{revtex4}
\usepackage{epsfig,graphicx}

\begin{document}

\title{\bf Elastic constants of $\beta$-eucryptite: A density functional theory study}

\author{Badri Narayanan,$^1$ Ivar E. Reimanis,$^1$ Edwin R. Fuller, Jr.,$^2$ and Cristian V. Ciobanu$^{3,*}$}
\affiliation{$^1$Department of Metallurgy \& Materials Engineering, Colorado School of Mines, Golden, Colorado 80401 \\
$^2$Ceramics Division, National Institute of Standards and Technology, Gaithersburg, Maryland 20899 \\
$^3$Division of Engineering, Colorado School of Mines, Golden, Colorado 80401}

\begin{abstract}
The five independent elastic constants of hexagonal $\beta$-eucryptite have been determined using density functional theory (DFT) total energy calculations. The calculated values agree well, to within 15\%, with the experimental data. Using the calculated elastic constants, the linear compressibility of $\beta$-eucryptite parallel to the $c$-axis, $\chi_c$, and perpendicular to it, $\chi_a$, have been evaluated. These values are in close agreement to those obtained from experimentally known elastic constants, but are in contradiction to the direct measurements based on a three-terminal technique. The calculated compressibility parallel to the $c$-axis was found to positive as opposed to the negative value obtained by direct measurements. We demonstrate that $\chi_c$ must be positive and discussed the implications of a positive $\chi_c$ in the context of explaining the negative bulk thermal expansion of $\beta$-eucryptite.
\end{abstract}

\maketitle

\section{Introduction}

Glass ceramics in $\mbox{Li}_2\mbox{O}$-$\mbox{Al}_{2}\mbox{O}_{3}$-$\mbox{SiO}_{2}$ (LAS) systems have attracted a lot of attention over the last several decades due to their low or even negative coefficient of thermal expansion (CTE), as well as due to their chemical and thermal stability.\cite{Bach,Palmer,Xu1999} This class of materials has been extensively commercialized owing to their exotic physical properties which makes them suitable for industrial applications (e.g., heat exchangers) which require dimensional stability and thermal shock resistance.\cite{Lichtenstein1998,Lichtenstein2000} They are also used in very specific applications like telescope mirror blanks, high precision optical devices and ring laser gyroscope.\cite{Palmer,Xu1999} The hexagonal $\beta$-eucryptite is a prominent member of this class of materials. It has a highly anisotropic CTE\cite{Xu1999} (i.e., $\alpha_a = $  7.26 $\times \mbox{10}^{-6}$ perpendicular to the $c$ axis,  $\alpha_c = $  -16.35 $\times \mbox{10}^{-6}$ parallel to the $c$ axis) which leads to a slightly negative crystallographic average (bulk) CTE. $\beta$-eucryptite undergoes a reversible order-disorder structural transition at $\sim$ 755 K.\cite{Guth} It exhibits one dimensional superionic conductivity of Li$^+$ ions along the $c$-axis which makes it a suitable electrolyte in Li based batteries.\cite{Schulz1980} Most of these unusual properties of $\beta$-eucryptite are, in part, related to its crystal structure.

\begin{figure}[htbp]
\begin{center}
\mbox{
\scalebox{0.1}
{
\includegraphics{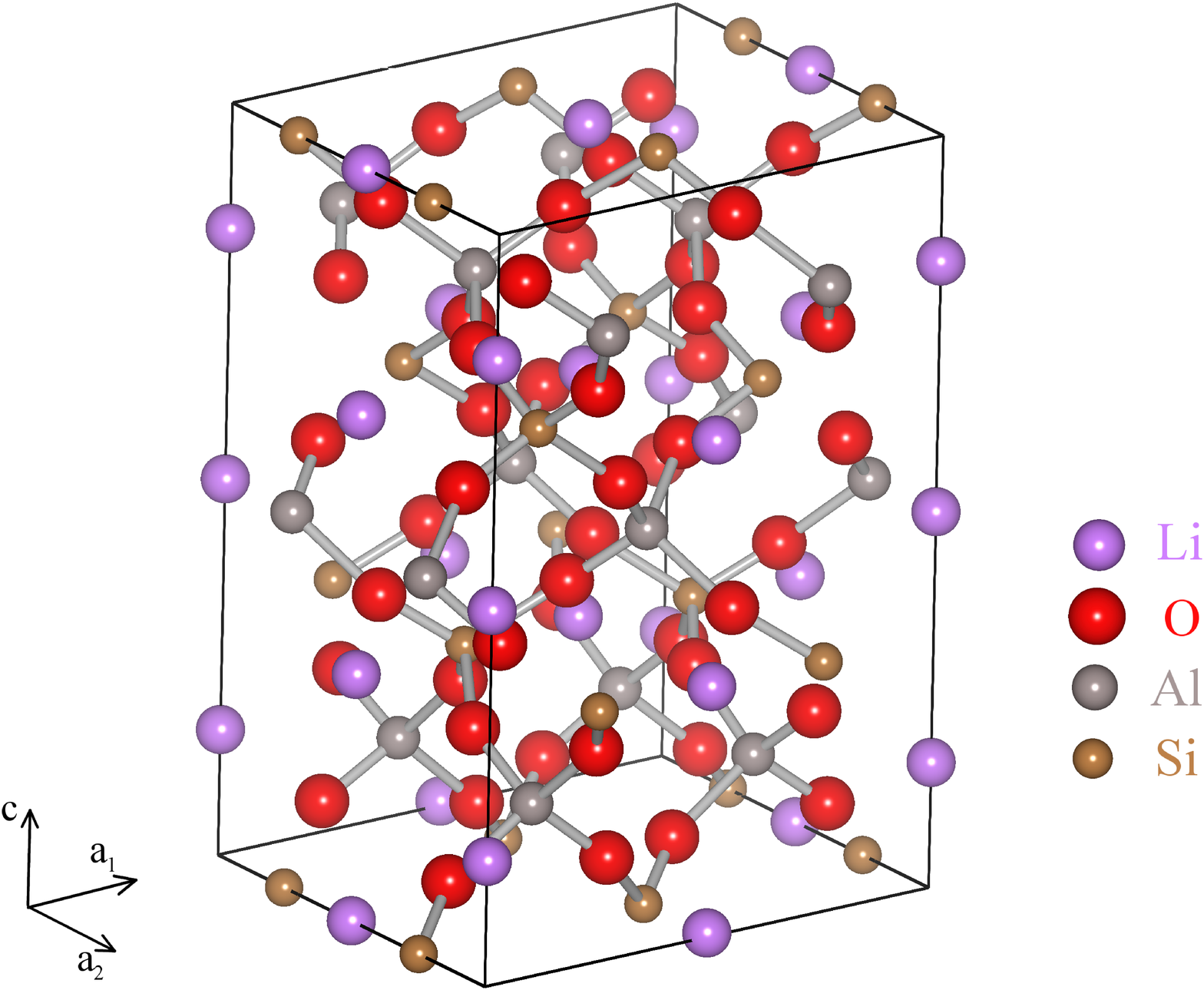}
}
}
\end{center}
\caption{(Color online) Crystal structure of ordered $\beta$-eucryptite containing 12 formula units of $\mbox{LiAlSiO}_{4}$ (84 atoms).}
\label{fig:Crystal}
\end{figure}

Figure \ref{fig:Crystal} illustrates a unit cell of $\beta$-eucryptite below the order-disorder transition containing 84 atoms with 12 unit formulae of LiAlSiO$_4$. A single crystal of ordered $\beta$-eucryptite, as shown by Figure \ref{fig:Crystal} has a primitive hexagonal structure belonging to the $P6_422$ space group.\cite{Winkler} This structure is a derivative of the $\beta$-quartz configuration, with half the Si$^{4+} $ ions replaced by Al$^{3+}$ while the charge imbalance is compensated by the channels of Li$^{+}$ ions parallel to the $c$-axis.\cite{Buerger} Several researchers\cite{Schulz_B28_1972,Tscherry_161_1972,Tscherry_175_1972,Pillars1973,Guth,Xu1999} have demonstrated through structural refinements that the structure is composed of interconnected helices of SiO${_4}^{4-}$ and AlO${_4}^{5-}$ tetrahedra with alternation of layers containing Si and Al atoms respectively, leading to a doubling of the $c$-axis of $\beta$-quartz.

As mentioned, the slightly negative crystallographic average CTE of $\beta$-eucryptite is due to the anisotropy of the linear expansion where a temperature increase involves a contraction along the $c$-axis which overcompensates the concomitant expansion in the plane perpendicular to the $c$-axis. Several theories have been proposed to explain this unusual thermal behavior.\cite{Gillery1959,Schulz1974,Moya1974,Xu1999} Hortal \emph{et al.}\cite{Hortal1975} employed a three-terminal technique\cite{3terminal} to measure the linear compressibility $\chi$ of $\beta$-eucryptite along the $a$ and $c$ axes, and the reported values of $\chi_a = (22.4 \pm 6.0)\times 10^{-3}$ GPa$^{-1}$  and $\chi_c = (-1.13 \pm 1.0) \times 10^{-3}$ GPa$^{-1}$. These measured values of $\chi$ supported the explanation given by Gillery and Bush\cite{Gillery1959} that the negative bulk thermal expansion is an elastic effect associated with the interconnected helices of Si and Al tetrahedra.

It is well known that linear compressibility along any direction in a crystal can be calculated from the elements of the stiffness matrix $C_{ij}$.\cite{Nye} Linear compressibilities can also be calculated from experimentally determined\cite{Haussuhl1984} stiffness constants; we carried out this calculation and the corresponding error analysis, and have found that $\chi_a = (2.57 \pm 0.02) \times 10^{-3}$ GPa$^{-1}$ and $\chi_c = (4.60 \pm 0.05) \times 10^{-3}$ GPa$^{-1}$ at a temperature of 293K. It should be noted that $\chi_c$ calculated using the experimentally known $C_{ij}$ is positive, in contrast to the negative value of $(-1.13\pm 1.0)\times 10^{-3}$ GPa determined by direct measurements.\cite{Hortal1975} While the sign of the direct measurement of $\chi_c$ remains in doubt due to experimental uncertainty, the near-zero value is also very different from that obtained in calculations. Since the sign of $\chi_c$ is linked with the explanation of negative CTE of $\beta$-eucryptite, it is necessary to address and possibly resolve the contradiction surrounding the sign of $\chi_c$.

In this paper, we compute the elastic stiffness constants of ordered $\beta$-eucryptite containing 84 atoms per unit cell in the framework of density functional theory (DFT). We then use the elastic constants to evaluate the linear compressibilities $\chi_a$ and $\chi_c$ to clarify the sign of linear compressibility parallel to the $c$-axis. Since the density functional theory (DFT) calculations offer an independent method of determining the linear compressibility, the present study can resolve the discrepancy discussed above. After ascertaining the sign of $\chi_c$, we discuss its implications on the explanation of negative crystallographic average CTE of $\beta$-eucryptite. We demonstrate that the negative CTE of $\beta$-eucryptite must arise from a combination of several interconnected phenomena as suggested by Xu \emph{et al.} \cite{Xu1999} and is related to a negative Gr\"{u}neissen function along the $c$-axis, rather than to the elastic effect proposed by Gillery and Bush.\cite{Gillery1959}

The paper is organized as follows: Sec II describes the methodology we adopted to calculate the elastic stiffness constants and the details of the DFT calculations; Sec III describes the results obtained in the present study which are discussed in Sec IV in the context of resolving the discrepancy and explaining the negative CTE of $\beta$-eucryptite; Sec V summarizes the results and describes our main conclusions.

\section{Methodology}

\subsection{Calculation of elastic constants}
In general, a crystal deforms in a homogeneous linear elastic manner when subjected to sufficiently small strains $\epsilon_{ij}$ ($i,j = 1,2,3$). The components $C_{ijkl}$ of the adiabatic stiffness matrix are the derivatives of elastic energy density with respect to the strain components:\cite{Wallace}

\begin{equation}
    C_{ijkl} =\frac{1}{V_0} \frac{{\partial}^{2}E}{\partial {\epsilon}_{ij}\partial{\epsilon}_{kl}}
\end{equation}
where $E$ is the elastic energy stored in a domain of volume $V$ of the crystal subjected to homogeneous deformations, and $V_0$ is the volume of the unstrained crystal.

In this section, we briefly describe the technique\cite{Fast1995} we employed to calculate the elastic constants of $\beta$-eucryptite. The lattice of hexagonal $\beta$-eucryptite is spanned by three primitive Bravais lattice vectors which can be written in a matrix form as:
\begin{equation}
\mathbf{R} = \left(
\begin{array}{ccc}
\frac{a}{2} & \frac{-\sqrt{3}a}{2} & 0 \\
\frac{a}{2} & \frac{\sqrt{3}a}{2} & 0 \\
0 & 0 & c
\end{array}
\right)
\end{equation}
 where each row is a lattice vector, and $a$, $c$ are the two lattice parameters that characterize the hexagonal structure. The vectors of the deformed lattice (${\mathbf{R}}^{\prime}$) can be obtained by multiplying $\mathbf{R}$ with a  distortion matrix $\mathbf{D}$:
\begin{equation}
\mathbf{R}^{\prime} = \mathbf{R}\mathbf{D},
\end{equation}
where $\mathbf{D}$ is defined in terms of the components of strain tensor as:
\begin{equation}
\label{distortion}\mathbf{D} = \left(
\begin{array}{ccc}
1+{\epsilon}_{11} & {\epsilon}_{12} & {\epsilon}_{13} \\
{\epsilon}_{21} & 1+{\epsilon}_{22} & {\epsilon}_{23} \\
{\epsilon}_{31} & {\epsilon}_{32} & 1+{\epsilon}_{33}
\end{array}
\right)
\end{equation}

The elastic energy $E$ of a crystal subjected to a general elastic strain ($\epsilon_{ij}$) can be expressed by means of a Taylor expansion in the distortion parameters truncated at the second order of strain.\cite{Wallace}
\begin{equation}
\label{Wallace_eq}
E(V,\mathbf{\epsilon}) = E_0 + V_0 \left(\sum_{i,j} \sigma_{ij} \delta_{ij}  + \sum_{i,j,k,l}\frac{1}{2}C_{ijkl}\epsilon_{ij}\epsilon_{kl}\right)
\end{equation}
where $E_0$ is the energy of a crystal volume $V_0$ at equilibrium, ${\sigma}_{ij}$ are the elements of the stress tensor, and $\delta_{ij}$ is the Kronecker symbol. Since the distortion matrix is symmetric, it is convenient to express Eq.~(\ref{Wallace_eq}) using the Voigt notation (11 = 1, 22 = 2, 33 = 3, 23 = 4, 31 = 5, and 12 = 6):
\begin{equation}
\label{Voigt}
E(V,\mathbf{\epsilon}) = E_0 + V_0 \left( \sum_{i} \sigma_{i} \epsilon_{i} \eta_{i} + \sum_{i,j}\frac{1}{2}C_{ij}\epsilon_{i}\eta_{i}\epsilon_{j}\eta_{j}\right)
\end{equation}
where $\eta_i =1 $ if  $i = 1, 2, \mbox{or } 3$ and $\eta_i= 2$ if $i = 4, 5, \mbox{or } 6$.

Due to the specific symmetry of the hexagonal lattice, there are only five {\em independent} elastic constants,\cite{Nye} which in Voigt notation are $C_{11}$, $C_{12}$, $C_{13}$, $C_{33}$ and $C_{44}$. These constants can be determined from specific distortion matrices for the hexagonal structures.\cite{Fast1995} Table~\ref{table:D-matrix} lists the distortion matrices used in the present study; for these matrices Eq.~(\ref{Voigt}) takes the simple form
\begin{equation} \label{eq:final}
E(V,\delta) = E_0 + V_0(A_1\delta + A_2{\delta}^{2}),
\end{equation}
where $A_1$ is related to stress components $\sigma_{ij}$, and $A_2$ is a linear combination of the elastic constants $C_{ij}$. The relationships between the second-order coefficient $A_2$ and the independent elastic constants for different strains are also given in Table \ref{table:D-matrix}.
For each of the five different deformations listed in Table \ref{table:D-matrix}, the total energy of the crystal was calculated for values of parameter $\delta$ varying from $-$5\% to $+$5\%. The zeroth, first, and second order coefficients in Eq.~(\ref{eq:final}) were extracted by means of polynomial fits of the total energy versus $\delta$ data. Using the relationships in the left column of Table \ref{table:D-matrix}, the elastic constants were extracted from the coefficients $A_2$ of the distortions considered. For all the cases, we found that the contribution of order 3 and higher terms to the energy in Eq.~(\ref{Wallace_eq}) was negligible, which confirms that the strains used are within the linear elastic limit.

\begin{table}
\caption{The distortion matrices and elastic constants for a hexagonal lattice.} \label{table:D-matrix}
\begin{center}
\begin{tabular}{cc} \toprule
Distortion matrix &  Second-order coefficient\\
		  &  $A_2$ in Eq.~(\ref{eq:final}) \\
\colrule
&           \\
$\left( \begin{array}{ccc} 1+\delta & 0 & 0 \\ 0 & 1+\delta & 0 \\ 0 & 0 & 1 \end{array} \right)$ &  $C_{11} + C_{12}$      \\
            &        \\
$\left( \begin{array}{ccc} 1+\delta & 0 & 0 \\ 0 & 1-\delta & 0 \\ 0 & 0 & 1 \end{array} \right)$ & $C_{11} - C_{12}$      \\
  &                  \\
$\left( \begin{array}{ccc} 1+\delta & 0 & 0 \\ 0 & 1+\delta & 0 \\ 0 & 0 & 1+\delta \end{array} \right)$ & $\begin{array}{c} C_{11} + C_{12}\\ \\ + 2C_{13} +\frac{C_{33}}{2}\end{array}$      \\
  &                  \\
$\left( \begin{array}{ccc} 1 & 0 & 0 \\ 0 & 1 & 0 \\ 0 & 0 & 1+\delta \end{array} \right)$ &  $C_{33}/2$      \\
  &                  \\
$\left( \begin{array}{ccc} 1 & 0 & \delta \\ 0 & 1 & 0\\ \delta & 0 & 1 \end{array} \right)$ & $2C_{44}$      \\
  &                 \\
\botrule
\end{tabular}
\end{center}
\end{table}

\subsection{Details of density functional calculations}
The total energy calculations for $\beta$-eucryptite supercells were performed within the framework of density functional theory (DFT) using the generalized gradient approximation (GGA) and the projector augmented wave (PAW) potentials\cite{PAW} implemented in the ab-initio simulation package VASP.\cite{VASP1,VASP2} We used the Perdew-Wang exchange-correlation function.\cite{PW91} The plane wave energy cutoff was set to 500 eV, and the augmentation charge cutoff to 605 eV. The computational cell consisted of one primitive cell of hexagonal $\beta$-eucryptite having 84 atoms, i.e. 12 formula units (f.u.) of LiAlSiO$_4$. The Brillouin zone was sampled with a 3$\times$3$\times$3 $\Gamma$-centered Monkhorst-Pack grid, which resulted in 14 irreducible $k$-points. The atomic coordinates were optimized using the conjugate gradient algorithm until the force components on any atom were smaller than 0.01 eV/\AA.

\section{Results}
\begin{figure}[htbp]
\begin{center}
\includegraphics[width=7.3cm]{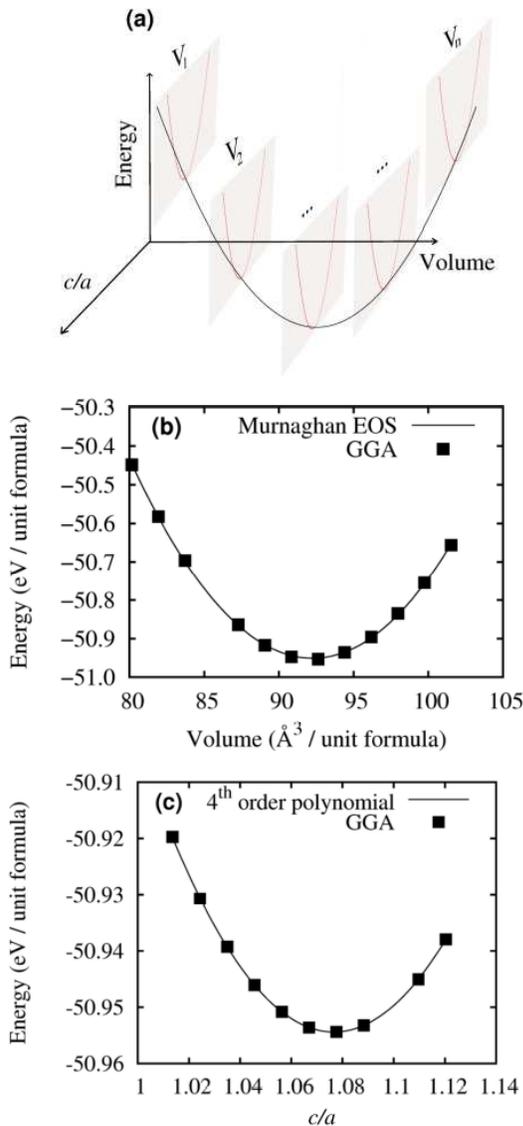}
\caption{Determination of lattice constants $a$ and $c$ from total energy GGA calculations. (a) Schematic representation of the procedure to obtain the structural parameters of $\beta$-eucryptite at equilibrium. (b) Energy vs. volume curve for the supercell shown in Fig.~\ref{fig:Crystal}. The black squares are the DFT calculated points, while the solid line represents the fit to the Murnaghan equation of state. (c) Energy as a function of the ratio $c/a$ at the equilibrium volume $V_0$ given by the Murnaghan fit. The squares are the DFT calculated values, and the solid line is a fourth-order polynomial fit.} \label{fig:EOS}
\end{center}
\end{figure}

To determine the equilibrium lattice parameters $a$ and $c$ of $\beta$-eucryptite, the ab-initio total energy calculations were performed for different values of the supercell volume $V$ and the $c/a$ ratio. The volume of the supercell was varied from $-15$\% to $+15$\% of the experimental value\cite{Pillars1973,Xu1999} by changing the parameter $a$ while keeping the ratio $c/a$ fixed. The procedure for finding the lattice constants $a$ and $c$ can be followed on the schematics in Fig.~\ref{fig:EOS}(a). At any given volume $V$ [which is a constant-volume plane in Fig.~\ref{fig:EOS}(a)], the supercell was relaxed and the total energy $E$ was computed for different $c/a$ values ranging from $1.00$ to $1.12$; the $E$ vs. $c/a$ data was then fitted with a fourth-order polynomial to determine the minimum energy for the given volume $V$. These minimum energy values found for different volumes $V$ are plotted in Fig.~\ref{fig:EOS}(b) and show an excellent fit to the Murnaghan equation of state\cite{Murnaghan1944}
 \begin{equation}
\label{Murnaghan_eq}
E(V) = E_0 + \frac{BV}{B^{\prime}}\left (\frac{{(V_0/V)}^{B^{\prime}}}{B^{\prime} - 1} +1 \right ) - \frac{BV_0}{B^{\prime} -1},
\end{equation}
where  $E_0$ is the energy corresponding to the equilibrium volume $V_0$,  $B$ is the bulk modulus at zero pressure, and ${B}^{\prime}=\left ({\partial B}/{\partial P}\right )_{T = 0}$ is the pressure derivative of the bulk modulus at 0 K. The parameters obtained from the fitting of $E$ versus $V$ data against the Murnaghan equation of state are listed in Table~\ref{table:Murnaghan}.

The bulk modulus obtained from the fit, as shown by Table \ref{table:Murnaghan}, is in excellent agreement with the experimentally determined value\cite{Haussuhl1984} which indicates a good agreement of the calculated $E$ versus $V$ data with Eq.~(\ref{Murnaghan_eq}). At the equilibrium volume $V_0$ obtained from the Eq.~(\ref{Murnaghan_eq}), the energy of the supercell was calculated for different $c/a$ values and plotted in Fig.~\ref{fig:EOS}(c). The $E$ vs. $c/a$ data at constant volume $V_0$ was fitted against a fourth-order polynomial to determine the optimum $c/a$ ratio at $V_0$. The optimized lattice constants $a$ and $c$ are subsequently extracted from the optimum $c/a$ ratio and the equilibrium volume $V_0$. The calculated values $a$ and $c$ (see Table~\ref{table:Murnaghan}) are in close agreement to the experimental ones $a_{exp}$ and $c_{exp}$, i.e., $a$ = 1.01${a}_{exp}$ and $c$ = 1.015${c}_{exp}$, which lead to $V$ = 1.03$\mbox{V}_{exp}$.

\begin{table}
\caption{Calculated lattice parameters, equilibrium volume $V_0$, bulk modulus $B$, and its pressure derivative (${B}^{\prime}=\left ({\partial B}/{\partial P}\right )_{T = 0}$) of $\beta$-eucryptite. The experimental values are also provided wherever available.} \label{table:Murnaghan}
\begin{center}
\begin{tabular}{lcccccccccc}
\toprule
&&&&&&&&&&\\
Technique & & ${a}$(\AA) & & ${c}$(\AA) & & ${V}_{0}$(${\mbox{\AA}}^{3}$/uf) & &$B$(GPa) && $B^{\prime}$ \\
\colrule
     &&        &&        &&        &&         &&       \\
GGA  && 10.594 && 11.388 && 92.25  && 102.27  && -1.05 \\
     &&        &&        &&        &&         &&       \\
Exp  && 10.497$^a$ && 11.200$^a$ && 89.06$^a$  && 109.9$^b$     &&       \\
     &&        &&        &&        &&         &&       \\
\botrule
$^a$\footnotesize{Ref.~\onlinecite{Pillars1973}, 293K.}\\
$^b$\footnotesize{From $C_{ij}$ in Ref.~\onlinecite{Haussuhl1984},}  \\
\footnotesize{extrapolated to 0 K.}
\end{tabular}
\end{center}
\end{table}

\begin{figure}
  \begin{center}
	\includegraphics[scale=0.05]{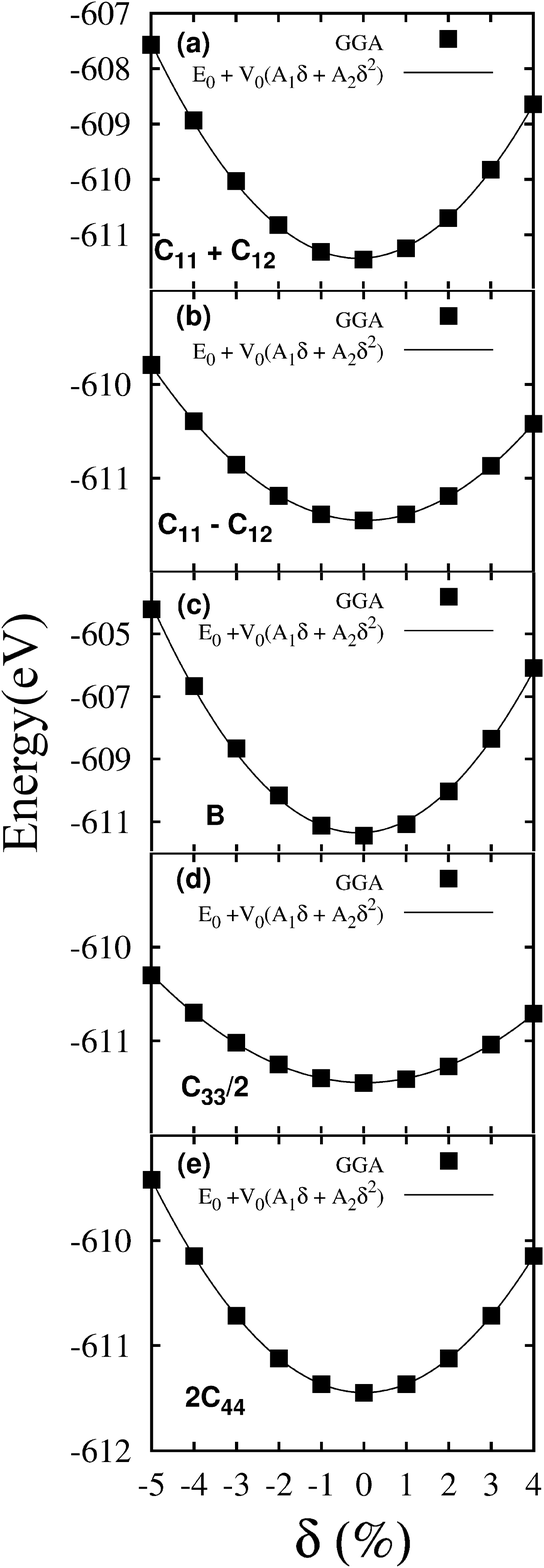}
\caption {Total energy of ordered $\beta$-eucryptite as a function of deformation parameter $\delta$ for the five different distortions (a) - (e) in the same order as listed in Table \ref{table:D-matrix}.}  \label{fig:Evsdelta}
\end{center}
\end{figure}

Using the calculated lattice parameters, we determined the five independent elastic constants of $\beta$-eucryptite at 0 K by employing the technique outlined in Section II. Figure \ref{fig:Evsdelta} shows the energy as a function of the deformation parameter $\delta$ for the different types of strain listed in Table~\ref{table:D-matrix} along with the corresponding fit polynomials. The elastic constants were evaluated from the second order coefficients $A_2$ of the fit polynomials through their relationships with the stiffness constants listed in Table~\ref{table:D-matrix}.

\begin{table}
\caption{Comparison of the calculated stiffness constants ${C}_{ij}$  of $\beta$-eucryptite with the experimental data from Ref.~\onlinecite{Haussuhl1984} extrapolated to 0 K using the thermoelastic constants $T_{ij} = d\log C_{ij}/dT$. The uncertainty
in any of the experimental values (Exp) is smaller than 2.5 GPa.} \label{table:elastic}
\begin{center}
\begin{tabular}{lcccccccccc}
\toprule
&&&&&&&&&&\\
          &&  $C_{11}$ &&  $C_{12}$ &&  $C_{13}$ &&  $C_{33}$ &&  $C_{44}$\\
\colrule
     &&        &&        &&       &&        &&          \\
GGA: (GPa)  && 165.64 && 70.98  && 78.59 && 132.83 && 58.68    \\
     &&        &&        &&       &&        &&          \\
Exp: (GPa) && 176.3 && 68.5  && 89.8 && 139.9 && 61.2    \\
     &&        &&        &&       &&        &&          \\
$T_{ij}$ $(10^{-3}/K)$ && -0.14 && 0.13 && -0.27 && -0.42 && -0.24   \\
     &&        &&        &&       &&        &&          \\
\botrule
\end{tabular}
\end{center}
\end{table}

The elastic constants have been measured experimentally by Hauss{\"u}hl \emph{et al.} using an ultrasonic technique at ambient temperature, 293K.\cite{Haussuhl1984} In order to compare at the same temperature the values of $C_{ij}$ computed in the present study (GGA) with the experiments, we have extrapolated the measured values of $C_{ij}$ to 0 K by using the thermoelastic constants $T_{ij} = d \log C_{ij}/dT$.\cite{Haussuhl1984} The calculated elastic constants that we obtained are within $\sim$15\% of the experimental values extrapolated to 0 K (see Table~\ref{table:elastic}). Furthermore, we have also found that the extrapolation of the GGA elastic constants to 273 K is also consistent with the experimental data at 273 K.

The linear compressibilities $\chi_a$ and $\chi_c$ (along the $a$ and $c$ axes) for a transversely isotropic material are related to the elastic constants $C_{ij}$ through:\cite{Nye}
\begin{eqnarray}
\label{eq:compress}
\chi_a &= &\frac{C_{33} - C_{13}}{C_{11}C_{33} - 2C_{13}^2 +C_{12}C_{33}} \nonumber  \\
\chi_c &= &\frac{C_{11} + C_{12}- 2C_{13}}{C_{11}C_{33} - 2C_{13}^2 +C_{12}C_{33}}
\end{eqnarray}

With the calculated elastic constants (Table~\ref{table:elastic}) in Eqs.~(\ref{eq:compress}), we determined the linear compressibilities of $\beta$-eucryptite at 0 K. However, the corresponding experimental data\cite{Hortal1975} obtained by direct measurements using a three-terminal method\cite{3terminal} have been reported at 273 K. In order to make comparisons of compressibility values at the same temperature
we have extrapolated the $C_{ij}$ values from GGA calculations to 273 K using the thermoelastic constants $T_{ij}$ from Ref.~\onlinecite{Haussuhl1984}. Similarly, we evaluated the experimentally determined elastic constants at 273 K. The extrapolated values of the elastic constants were then used in Eq.~(\ref{eq:compress}) to determine $\chi_a$ and $\chi_c$, which were compared with the direct measurements (Table~\ref{table:compressibility}).
We should note that the uncertainty in the GGA values for compressibility comes solely from propagating the uncertainties in the thermoelastic constants.\cite{Haussuhl1984}

\begin{table}
\caption{Comparison of calculated values of linear compressibility of ordered $\beta$-eucryptite $\chi_a$  and $\chi_c$ with experimental data. The GGA elastic constants and the experimental values\cite{Haussuhl1984} have been extrapolated to 273 K to calculate $\chi_a$ and $\chi_c$. The calculated values are in good agreement with each other. However, they are in contradiction to the direct measurements\cite{Hortal1975} at 273 K.}
\label{table:compressibility}
\begin{center}
\begin{tabular}{c c c c}
\toprule
&&& \\
Parameter & GGA  & Ref. \onlinecite{Haussuhl1984} & Ref. \onlinecite{Hortal1975}\\
&$(10^{-3}$ GPa$^{-1})$ & $(10^{-3}$ GPa$^{-1})$ & $(10^{-3}$ GPa$^{-1})$ \\
\colrule
&  &   & \\
$\chi_a$ & 2.67 $\pm$ 0.06 & 2.58 $\pm$ 0.02   & 22.4 $\pm$ 6.0  \\
  &            &      &\\
$\chi_c$ & 5.20 $\pm$ 0.15 &  4.52$\pm$ 0.05  & -1.13 $\pm$ 1.0 \\
  & 	      &        & \\
\botrule
\end{tabular}
\end{center}
\end{table}

\section{Discussion}
 Our calculated linear compressibility values of ordered $\beta$-eucryptite, extrapolated to 273 K, agree well with those derived using experimentally determined stiffness constants.\cite{Haussuhl1984} These two sets of compressibility values, however, are in contradiction with the direct measurements\cite{3terminal} reported by Hortal \emph{et al.}\cite{Hortal1975} We note that the calculated values of $\chi_c$ are positive, while the value reported from direct measurements is negative (refer to Table~\ref{table:compressibility}). Furthermore, the measured value of $\chi_a$ is about one order of magnitude larger than that calculated in the present study.

We now focus on the implications of the calculated compressibility values on the thermal behavior of $\beta$-eucryptite, in particular on the coefficient of thermal expansion. An early study by Munn\cite{Munn1972} addresses the effect of anisotropy of elastic properties on the thermal expansion in the quasi-harmonic approximation, where the the vibrations are taken to be harmonic but with deformation-dependent frequencies. The bulk thermal expansion coefficient $\alpha$ of a hexagonal crystal can be written in terms of the thermal expansion coefficients along the $c$-axis ($\alpha_c$) and $a$-axis ($\alpha_a$) as:\cite{Munn1972}
\begin{equation}
\label{eq:CTE}
\alpha \equiv \frac{2\alpha_a+\alpha_c}{3} = \frac{H_t}{3V_0} (2\chi_a\gamma_a + \chi_c\gamma_c)
\end{equation}
where $H_t$ is the heat capacity at constant stress, and $\gamma_{a,c}$ are the Gr\"{u}neisen functions which describe the dependence of entropy on strain.\cite{Gruneisen1912}

Using structural arguments, Moya \emph{et al.} have asserted that both $\chi_a$ and $\gamma_a$ must be positive.\cite{Moya1974} This assertion in combination with Eq.~(\ref{eq:CTE}) suggests that the bulk thermal expansion coefficient $\alpha$ can be negative only if $\chi_c$ or $\gamma_c$ is negative but not both. Early measurements\cite{Moya1974} yielded a negative value for $\chi_c$, which implied that the bulk thermal expansion coefficient of $\beta$-eucryptite was negative because of the negative compressibility along the $c$-axis.
Hortal \emph{et al.} \cite{Hortal1975} put forth the idea that a negative $\chi_c$ would explain the negative bulk thermal expansion in $\beta$-eucryptite as an elastic effect associated with the interconnected Si and Al-tetrahedra as proposed by Gillery and Bush.\cite{Gillery1959}

Our results are, however, in contradiction to this point of view. The value we calculated for $\chi_c$ using the elastic constants turns out to be positive. According to Eq.~(\ref{eq:CTE}), a positive value for $\chi_c$ implies that the Gr\"{u}neisen function $\gamma_c$ must be negative in order to obtain a negative bulk expansion coefficient $\alpha$. Indeed, recent phonon spectra calculations by Lichtenstein \emph{et al.}\cite{Lichtenstein1998,Lichtenstein2000} show that the Gr\"{u}neisen parameters parallel to the $c$-axis for the modes around 400 cm$^{-1}$ (bending of the Al-O and Si-O bonds) are large and negative, which leads to a negative value of $\gamma_c$. Lichtenstein and coworkers attributed the negative $\gamma_c$ to Li-position disordering and proposed an explanation for negative bulk CTE of $\beta$-eucryptite similar to that of Schulz.\cite{Schulz1974} Independently, Xu \emph{et al.} used powder synchotron X-ray and neutron diffraction to show that cation disordering alters the structure of $\beta$-eucryptite and significantly affects its thermal behaviour.\cite{Xu1999} They have shown that Al/Si and Li disorder leads to a significant decrease in the lattice parameter $c$ with only a moderate increase in $a$, leading to an overall volume contraction of $\sim1$\%.This behavior was explained\cite{Xu1999} as a combined effect of several interconnected phenomena including tetrahedral tilting, tetrahedra flattening, and shortening of the Si-O and Al-O bonds.

Thus, our results are consistent with the findings of Lichtenstein \emph{et al.}\cite{Lichtenstein1998,Lichtenstein2000} and Xu \emph{et al.}\cite{Xu1999} leading us to conclude that the negative coefficient of thermal expansion of $\beta$-eucryptite is due to a negative value of $\gamma_c$ associated with cation disordering, rather than to a negative $\chi_c$ as proposed by Hortal \emph{et al}.\cite{Hortal1975}

\section{Conclusion}
To summarize, we have computed the elastic stiffness constants of ordered $\beta$-eucryptite containing 84 atoms per unit cell within the framework of generalized gradient approximation of DFT. The calculated elastic constants are in close agreement with the experimentally known values. The elastic constants were subsequently used to compute the linear compressibilities of $\beta$-eucryptite parallel and perpendicular to the $c$-axis. Our calculated compressibility values agree well with those calculated from experimentally known elastic constants as reported by Hauss{\"u}hl \emph{et al.}\cite{Haussuhl1984} The calculated values of compressibility are, however, in contradiction to those reported by Hortal \emph{et. al} who measured the compressibilities $\chi_c$ and $\chi_a$ using a direct three-terminal method. Our calculations show that the compressibility parallel to the $c$-axis is positive as opposed to the negative value obtained from the direct measurements.\cite{Hortal1975}

Based on our calculations, we have also shown that the negative bulk thermal expansion of $\beta$-eucryptite must be associated with a negative Gr\"{u}neissen function parallel to the $c$-axis rather than with a negative compressibility as proposed by Hortal \emph{et al.}. The conclusion that the negative bulk thermal expansion coefficient occurs because of a negative Gr\"{u}neissen function is consistent with the results of Lichtenstein \emph{et al.},\cite{Lichtenstein2000} who showed through the calculations of phonon density of states that the Gr\"{u}neissen function parallel to the $c$-axis is strongly negative due to the ``bending" modes of the Si-O and Al-O bonds. Our results are also consistent with the neutron diffraction and X-ray synchrotron diffraction studies conducted by Xu \emph{et. al}.\cite{Xu1999}

The present study in conjunction with the results of Lichtenstein \emph{et al.} and Xu \emph{et al.} clearly indicates that the $\chi_c$ must be positive and that the negative bulk thermal expansion is due to to cation disordering,\cite{Xu1999} rather than to elastic effects.\cite{Gillery1959}

\begin{acknowledgments}
The authors gratefully acknowledge financial support from the Department of Energy's Office of Basic Energy Sciences through Grant No. DE-FG02-07ER46397 (B.N. and I.E.R.) and from the National Science Foundation through Grant No CMMI-0846858 (C.V.C.). The supercomputing resources were provided by the Golden Energy Computing Organization at Colorado School of Mines acquired with financial assistance from the NSF (Grant No. CNS-0722415) and the National Renewable Energy Laboratory.
\end{acknowledgments}

\end{document}